\documentclass[aps,twocolumn,showpacs,preprintnumbers,nofootinbib,prd,superscriptaddress,10pt]{revtex4-1}

\makeatletter
\def\l@subsubsection#1#2{}
\def\l@subsubsubsection#1#2{}
\makeatother

\usepackage{comment}
\usepackage{graphicx,amssymb,amsmath,amsthm,amsfonts,epsfig,epsf,fixmath}
\usepackage[usenames]{color}
\usepackage{xcolor}
\usepackage{epstopdf}
\usepackage{amsmath}
\usepackage{float}

\usepackage{aas_macros}
\usepackage{bm}
\usepackage{dcolumn}
\usepackage{lipsum}
\usepackage{latexsym}
\usepackage{rotating}
\usepackage{longtable}

\usepackage{enumerate}
\usepackage{tensor,multirow}
\usepackage{url}
\usepackage[linktocpage]{hyperref}

\begin{document}

\title{Dark photon superradiance quenched by dark matter}

\author{Enrico Cannizzaro}
\affiliation{Dipartimento di Fisica, ``Sapienza'' Universit\`a di Roma \& Sezione INFN Roma1, Piazzale Aldo Moro
5, 00185, Roma, Italy}
\author{Laura Sberna}
\affiliation{Max Planck Institute for Gravitational Physics (Albert Einstein Institute) Am Mu\"{u}hlenberg 1, 14476 Potsdam, Germany}
\author{Andrea Caputo}
\affiliation{School of Physics and Astronomy, Tel-Aviv University, Tel-Aviv 69978, Israel}
\affiliation{Department of Particle Physics and Astrophysics, Weizmann Institute of Science, Rehovot 7610001, Israel}
\author{Paolo Pani}
\affiliation{Dipartimento di Fisica, ``Sapienza'' Universit\`a di Roma \& Sezione INFN Roma1, Piazzale Aldo Moro
5, 00185, Roma, Italy}

\begin{abstract} 
Black-hole superradiance has been used to place very strong bounds on a variety of models of ultralight bosons such as axions, new light scalars, and dark photons. It is common lore to believe that superradiance bounds are broadly model independent and therefore pretty robust. In this work we show however that superradiance bounds on dark photons can be challenged by simple, compelling extensions of the minimal model. In particular, if the dark photon populates a larger dark sector and couples to dark fermions playing the role of dark matter, then superradiance bounds can easily be circumvented, depending on the mass and (dark) charge of the dark matter.
\end{abstract}

\maketitle

\section{Introduction}
It is well known that bosonic waves scattering off spinning black holes~(BHs) can extract rotational energy via a phenomenon called superradiance~\cite{1971ZhPmR..14..270Z} (see~\cite{Brito:2015oca} for an overview). This process takes place as long as $\omega< m\Omega_H$, where $\omega$ is the frequency of the wave, $m$ is its azimuthal number and $\Omega_H$ is the angular velocity of the BH horizon. If superradiant scattering is supported by a confinement mechanism of the modes, the extraction of energy happens at a continuous level, leading to the so-called superradiant instability~\cite{Brito:2015oca}. Remarkably, the bare mass of the bosonic field can serve for such purpose, as it can naturally confine low-frequency modes in the vicinity of the BH~\cite{Damour:1976kh, Detweiler:1977gy, Cardoso:2004nk}.

For the process to be efficient, the Compton wavelength of the modes must be roughly comparable with the size of the BH. For astrophysical BHs this corresponds to bosonic masses in the range $m_b\sim (10^{-21}-10^{-10}) \, \rm eV$. In this case, a macroscopic bosonic condensate can form around a spinning BH, leading to striking observable signatures such as gaps in the BH spin-mass (``Regge") plane and nearly monochromatic gravitational-wave emissions from the condensate~\cite{Arvanitaki:2009fg, Arvanitaki:2010sy}, leading to a rich phenomenology in isolated and binary systems~\cite{Brito:2015oca}. Hence, BH superradiance represents a powerful tool to probe exotic ultralight particles beyond the Standard Model, such as axions or dark photons~(DPs). 

Until recently, studies of the superradiant instability assumed that the superradiant field was free from interactions, as expected for a field only minimally coupled to gravity. However, as number densities can reach extreme values in the process, the effect of interactions can be crucial, even for very weakly interacting fields.
Recent studies have considered the effect of self-interactions, both for scalar and vector fields~\cite{Baryakhtar:2020gao, Clough:2022ygm}, axion-photon couplings~\cite{Ikeda:2018nhb, Boskovic:2018lkj}, interactions with astrophysical plasmas~\cite{Pani:2013hpa,Conlon:2017hhi, Blas:2020kaa, Dima:2020rzg, Cannizzaro:2020uap, Cannizzaro:2021zbp, Wang:2022hra}, and models of DPs kinetically mixed with Standard Model photons~\cite{Caputo:2021efm}. 

In this work, we consider the interaction of a (vector) superradiant field with other (fermionic) fields in the dark sector, which constitute the entirety or just a fraction of the dark matter~(DM). 
In other words, we consider an extended dark sector, populated not only by a massive DP, but also by dark fermions. 

The new dark fermions constitute a dark plasma, which can alter the dispersion relation of the DP and possibly affect its superradiant instability. Intuitively, the presence of a dark fermion should generate a gap in the dispersion relation of the DPs, effectively endowing it with a plasma mass 
\begin{equation}
\label{eq:plasmafreq}
\omega_{\rm pl}^\chi= \Big(\frac{4 \pi \alpha_\chi \rho_\chi}{m_\chi^2}\Big)^{1/2}, 
\end{equation}
where $\rho_\chi$ is the energy density of the dark fermions, $m_\chi$ its mass, and $\alpha_\chi=q_\chi^2/(4\pi)$ the fine structure constant within the dark sector. 

We assess the effect of an extended dark sector on the superradiant instability by studying quasibound states around nonspinning BHs. We solve numerically for the quasibound states of a DP in the presence of a dark plasma and we find -- as expected -- that the interaction with a (sufficiently dense) dark plasma can significantly alter the lifetime of quasibound states. Extrapolated to spinning BHs, our results indicate that superradiant bounds on DPs can be completely invalidated in motivated models granting a DM candidate. In particular, we will show that simple models of $\sim \rm TeV$ self-interacting DM~(SIDM)~\cite{Ackerman:2008kmp}, can generate $\omega_{\rm pl}^\chi \simeq 10^{-13}-10^{-14} \, \rm eV$ around the BH and make the superradiance timescale much longer than other astrophysical timescales, such as the accretion one, thus invalidating some of the current DP bounds.   

This work is organized as follows: in Sec.~\ref{sec:setup} we introduce the generic formalism to study linear perturbations of a massive spin-1 field in a plasma, recasting the set of differential equations into one master equation for the DP field. Next, in Sec.~\ref{sub:general equation} we specialise to Schwarzschild spacetime and write the corresponding set of differential equations. In Sec.~\ref{sec:results} we solve the field equations numerically and show how the quasibound spectrum of the DP field is altered by the presence of dark fermions. Then, in Sec.~\ref{sec:DM} we introduce an example of DM model and deduce a realistic estimate of dark plasma frequency sufficient to alter the spectrum. In Sec.~~\ref{sec:bound} we discuss the impact of our findings on the current DP bounds from BH superradiance. Finally, we conclude in Sec.~\ref{sec:discussion}.

Henceforth we use natural units and also impose $G = 1$. This is the reason why both a product like $M\mu$, where $M$ is the BH mass and $\mu$ the DP mass, and a ratio like $\omega_{\rm pl}^\chi/\mu$ will be dimensionless.\footnote{The reader more familiar with geometrized units, $G=c=1$, might find it useful to notice that $\hbar$ only appears in terms like $\hbar\omega_{\rm pl}^\chi$ and $\hbar\mu$, which have the dimensions of a mass. On the other hand, the reader more familiar with natural units might find it useful to notice that $G$ appears only in the Schwarzschild radius $r_H=2GM$ and in the dimensionless coupling $GM\mu$.}

\section{Setup}
\label{sec:setup}
In the following, we will consider a massive spin-1 field coupled to a dark fermion current. The dark sector is then described by the Lagrangian 
\begin{equation}
    \mathcal{L}=-\frac{1}{4}F_{\mu\nu}F^{\mu\nu}-\frac{1}{2}\mu^2 V_\mu V^\mu -J_\mu V^\mu \, ,
\end{equation}
where $V^\nu$ is the DP field,  $F^{\mu\nu}=\nabla^\mu V^\nu-\nabla^\nu V^\mu$ is the field strength, $\mu$ is the DP mass and $J^{\mu}$ is the dark sector current. In this work, we assume that the dark sector is secluded from the Standard Model. For instance, we assume that the kinetic mixing between dark and ordinary photons is sufficiently small that we can neglect it.


Varying the action leads to a Proca equation sourced by the dark current:
\begin{equation}
\label{eq:ProcaBackground}
    \nabla_\mu F^{\nu\mu}+\mu^2 V^\nu =q_\chi n u^\nu +J_{2}^\nu \, ,
\end{equation}
where $n$ is the density of the fermions, $q_\chi$ is their dark charge, $u^\mu$ is their four-velocity, and $J_2^\nu$ is the current of a second species which we assume to be present to neutralize the plasma. Note that deriving Eq.~\eqref{eq:ProcaBackground} and using the conservation of the currents leads to the Lorenz condition $\nabla_\mu V^\mu=0$.
The Proca equation must be solved jointly with the momentum equation describing the motion of the dark fermions 
\begin{equation}
\label{eq:momentumeq}
    u^{\mu}\nabla_{\mu}u^\nu=\frac{q_\chi}{m_\chi}F^{\nu\mu}u_{\mu} \, ,
\end{equation}
where $m_\chi$ is the fermion mass.
The conservation of the current also implies the continuity equation 
\begin{equation}
\label{eq:continuityeq}
    \nabla_\mu (n u^\mu) =0 \, .
\end{equation}

We will solve the system perturbatively by considering small perturbations for the DP field, the density and four velocity of the dark plasma (i.e. $F_{\mu\nu}=F_{\mu\nu}^{\rm background}+\Tilde{F}_{\mu\nu}$ and the same for $n$ and $u^\mu$). In order to simplify the problem, we neglect the perturbations in the second, oppositely charged species, $\Tilde{J_2}^\mu=0$, in analogy to the standard case of an electron-ion plasma. The presence of a second fermion would only shift the (dark) plasma frequency, with the exact amount depending on its mass and background density. Given the uncertainties in other parameters, we can re-absorb this shift in the following definition of the plasma frequency for one species.\footnote{For the vanilla DM model we will discuss in Section~\ref{sec:DM}, the situation is actually very similar to that of an electron-positron plasma, with the two species of opposite charges having the same mass~\cite{Ackerman:2008kmp}. In this case the change in the dark plasma frequency should roughly be a factor $\sim \sqrt{2}$~\cite{Stenson2017DebyeLA}.}
We will also neglect higher order perturbations and the backreaction of the field on the metric, as they are negligible, at least during the first stages of the superradiant instability.
The perturbed equations of motion are 
\begin{align}
\label{eq:perturbedprocaeq}
    \nabla_\mu \tilde{F}^{\nu\mu}+\mu^2 \tilde{V}^\nu&=q_\chi\tilde{n}u^\nu +  q_\chi n\tilde{u}^\nu \, ,\\
\label{eq:perturbedmomentumeq}
    \tilde{u}^\mu \nabla_\mu u^\nu+u^\mu \nabla_\mu \tilde{u}^\nu&=\frac{q_\chi}{m_\chi}\tilde{F}^{\nu\mu}u_\mu+\frac{q_\chi}{m_\chi} F^{\nu\mu}\tilde{u}_\mu \, ,\\
\label{eq:perturbednorm}
    \tilde{u}^\mu u_\mu&=0 \, , \\
\label{eq perturbedlorentz}
    \nabla_\mu \tilde{V}^\mu&=0 \, .
\end{align}

Following the procedure outlined in Refs.~\cite{1981A&A....96..293B,Cannizzaro:2020uap}, this set of equations can be reassembled into a third-order, differential master equation. We report the details of this procedure applied to our system in Appendix~\ref{app:Fieldeq}. The master equation for the linear perturbations of the DP field in the presence of a moving, magnetised (or not) plasma reads
\begin{align}
\label{eq:finaleq}
   &\nonumber h^\xi_{\ \alpha}u^\mu \nabla_\mu (\nabla_\sigma \tilde{F}^{\alpha \sigma }+\mu^2 \tilde{V}^\alpha)\\
   &\nonumber+(\theta_{\ \alpha}^{\xi}+\omega_{\ \alpha}^{\xi}+\theta h^{\xi}_{\ \alpha}+\omega^{\ \xi}_{L \ \ \alpha} )(\nabla_\sigma \tilde{F}^{\alpha \sigma }+\mu^2 \tilde{V}^\alpha)\\
      &+\frac{q_\chi}{m_\chi}E^\xi u_\alpha (\nabla_\sigma \tilde{F}^{\alpha \sigma }+\mu^2 \tilde{V}^\alpha)=  \omega_{\rm pl}^{\chi \, 2} \tilde{F}^{\xi\mu}u_\mu \, ,
\end{align}
where $\omega_{\rm pl}^\chi$ is the dark plasma frequency defined in Eq.~\eqref{eq:plasmafreq}, while $E^\alpha, \omega_L^{\alpha\beta}, \omega^{\alpha \beta}$ and $\theta^{\alpha \beta}$ are the background electric field, Larmor tensor, vorticity and deformation defined in Appendix~\ref{app:Fieldeq}. As anticipated, both the bare DP mass and the ``effective" plasma mass~\eqref{eq:plasmafreq} appear in this equation. It is straightforward but important to verify that
\begin{itemize}
    \item in the $\omega_{\rm pl}^\chi\rightarrow 0$ limit, Eq.~\eqref{eq:finaleq} reduces to the vacuum Proca equation,
\begin{equation}
    \nabla_\sigma \tilde{F}^{\sigma \alpha}=\mu^2 \tilde{V}^\alpha \, ;
\end{equation}
\item in the $\mu\rightarrow 0$ limit, Eq.~\eqref{eq:finaleq} matches the one for the Standard Model photon in a cold plasmic medium on an arbitrary spacetime background~\cite{1981A&A....96..293B}, once we identify the field and the plasma with the Standard Model ones.
\end{itemize}
Also, it is important to notice that Eq.~\eqref{eq:finaleq} is a third-order differential equation, at variance with the vacuum Proca case, and that it depends on the background plasma configuration.

\section{Perturbations on a Schwarzschild spacetime}
\label{sub:general equation}
We now specialize to the Schwarzschild background. We work in the standard coordinates $(t,r, \theta,\phi)$, in which the line element reads
\begin{equation}
    ds^2 = - f dt^2 + f^{-1} dr^2 + r^2 d\Omega_2^2 \, ,
\end{equation}
with $f(r) = 1- 2 M /r$, where $M$ is the BH mass.

Assuming that the background plasma is also spherically symmetric, it is convenient 
perform a multipolar expansion of the dark electromagnetic field as~\cite{Dolan:2012yt}
\begin{equation}
\label{eq:multexp}
    \tilde V_\mu(r,t,\theta,\phi)=\frac{1}{r}\sum_{i=1}^4\sum_{l,m}c_iu_{(i)}^{lm}(t,r)Z_\mu^{(i)lm}(\theta,\phi),
\end{equation}
where $Z_\mu^{(i)lm} $ are the vector spherical harmonics (which satisfy as usual orthogonality conditions) and $c_1=c_2=1$, $c_3=c_4=1/\sqrt{l(l+1)}$. This allows separating the angular dependence of the field from the radial dependence.

The behavior of the DP perturbations depends on the plasma profile through the dark plasma frequency $\omega^\chi_{\rm pl}$. In the following, we consider two different configurations: a static plasma, and a plasma in free fall.
A static plasma is not a realistic configuration, especially in the vicinity of the BH horizon, but its perturbation equations take a simple form, allowing us to understand the interplay between bare and effective masses analytically. Studying two plasma configurations will also allow us to explore the dependence of the quasibound states on the background plasma four-velocity.

\subsubsection{Static plasma}

Using the decomposition~\eqref{eq:multexp} and working in the frequency domain, $u_{(i)}^{lm}(t,r)=u_{(i)}^{lm}(r)e^{-i\omega t}$, we obtain a set of three equations from the radial and angular components of Eq.~\eqref{eq:finaleq}. Using the Lorenz condition, it is possible to close the system and rewrite the field equations in a more straightforward way. We give here the final form of the perturbation equations, and a detailed derivation in Appendix~\ref{app:Staticeqs}. 
Introducing the differential operator $\mathcal{D}_2 \equiv   \frac{d^2}{dr_*^{2}}+ \omega^2- f \Big(\frac{l(l+1)}{r^2}+\mu^2\Big)$, the equations read
\begin{align}
\label{eq:D2eq}
    \mathcal{D}_2 \ u_{(2)}=&\frac{1}{r^3(\omega^2-f \omega_{\rm pl}^{\chi \, 2})}f((2(-3M+r)\omega^2\\&\nonumber+f(6M+r(-2+l(l+1)+r^2\mu^2)) \omega_{\rm pl}^{\chi \, 2})u_{(2)}\\&\nonumber+2(3M-r)(\omega^2-f \omega_{\rm pl}^{\chi \, 2})u_{(3)}-f^2 r^2  \omega_{\rm pl}^{\chi \, 2} u_{(3)}' \, , \\
     \mathcal{D}_2 \ u_{(3)}=&\frac{f}{r^2\omega^2}(-l(l+1)(2\omega^2- f \omega_{\rm pl}^{\chi \, 2})u_{(2)}\label{eq:D3eq}\\&\nonumber+ \omega_{\rm pl}^{\chi \, 2}((r^2\omega^2-f l(l+1))u_{(3)}+f l(l+1)r u_{(2)}') ,  \\
    \mathcal{D}_2 \ u_{(4)}=& f \omega_{\rm pl}^{\chi \, 2} u_{(4)} \, . \label{eq:D4eq}
\end{align}
 Owing to the spherical symmetry of the background, the axial sector ($u_{(4)}$) is decoupled from the polar sector ($u_{(2)}$, $u_{(3)}$), the equations do not depend on the azimuthal number $m$, and there is no mixing between modes with different quantum number $l$. In the limit $\omega_{\rm pl}^\chi\rightarrow 0$, Eqs.~\eqref{eq:D2eq}-\eqref{eq:D4eq} reduce to the standard equations for a noninteracting Proca field in a Schwarzschild spacetime~\cite{Rosa:2011my}.

From Eqs.~\eqref{eq:D2eq}-\eqref{eq:D4eq}, it is immediate to see that the interplay between the bare mass and the plasma frequency is nontrivial. Naively, one might expect the perturbations to depends on a total mass (squared) given by the squared sum of the bare and effective masses, at least for transverse modes. However, this is true only for the axial sector, see Eq.~\eqref{eq:D4eq}. For the polar sector, the interplay is less straightforward. 

In the limit of flat spacetime, Eqs.~\eqref{eq:D3eq}-\eqref{eq:D4eq} read
\begin{equation}
    \Big(\frac{d^2}{dr_*^{2}}+\omega^2-\mu^2\Big)u_{(3,4)}= \omega_{\rm pl}^{\chi \, 2} u_{(3,4)} \, , 
\label{eq:axialboundaryinf}
\end{equation}
leading, in momentum space, to the dispersion relation
\begin{equation}
    \omega^2=k^2+(\mu^2+ \omega_{\rm pl}^{\chi \, 2}) 
\end{equation}
of transverse massive modes in a plasma, where the bare and effective masses are squared-summed. For the radial component of the field we instead obtain  
\begin{equation}
    \epsilon_{\rm pl}\Big(\frac{d^2}{dr_*^{2}}+\omega^2-\mu^2\Big)u_{(2)}=\frac{ \omega_{\rm pl}^{\chi\, 2}}{\omega^2} \mu^2 u_{(2)} \, ,
\end{equation}
where we introduced the plasma dielectric tensor $\epsilon_{\rm pl}=1- \omega_{\rm pl}^{\chi \, 2}/\omega^2$.
From this equation we can verify that our formalism recovers the expected phenomenology in two important limits:  
\begin{itemize}
    \item if $\mu\rightarrow 0$, the right-hand side vanishes and the only solution is $\epsilon_{\rm pl}=0$, i.e. $\omega^2= \omega_{\rm pl}^{\chi \, 2}$. In the absence of a mass, this degree of freedom does therefore become electrostatic. In fact, a massless spin-1 particle in a cold plasmic medium only propagates two degrees of freedom, while the third one is electrostatic~\cite{Raffelt:1996wa}.
    \item if $\omega_{\rm pl}^\chi\rightarrow 0$, $\epsilon_{\rm pl}\rightarrow 1$ we recover the dispersion relation of a propagating Proca degree of freedom in vacuum $\omega^2=k^2+\mu^2$. 
    For realistic plasma density profiles that vanish at spatial infinity -- as the one we will consider in this work (see Sec.~\ref{sec:densityprofile}) -- this degree of freedom behaves as a propagating Proca degree of freedom at large $r$.  
\end{itemize}

\subsubsection{Free-fall plasma}
Up to now we have modelled the plasma surrounding the BH as static~\cite{Cannizzaro:2020uap,Cannizzaro:2021zbp}. We now want to relax this approximation, by considering a more realistic free-falling plasma. A freely-falling massive particle follows the geodesics of the Schwarzschild spacetime, and its four velocity reads 
\begin{equation}
\label{eq:FFvelocity}
    u^{\mu}=((1-2M/r)^{-1},- \sqrt{2M /r}, 0, 0) \, .
\end{equation}
As its motion is purely radial, this plasma configuration does not break the spherical symmetry and therefore, when the field is decomposed in spherical harmonics, it does not generate couplings between different $l,m$ modes and the axial and polar sectors decouple as in the static-plasma case. 
Moreover, as the plasma four-velocity does not depend on time, the system is still stationary (at least at the linearized level when backreaction is neglected). Therefore, even in the case of a free-falling plasma, we can work in the frequency domain and perform the same multipolar expansion as in Eq.~\eqref{eq:multexp} with the assumption of a harmonic time dependence $\sim e^{-i\omega t}$. 
The corresponding field equations are much more involved than in the static plasma case, and are reported in Appendix~\ref{app:FFeq}. 

In the case of a static plasma, the field equations reduced to second order differential equations. In the case of free-fall, the perturbation equations remain of third order, and thus require three boundary conditions, as we will see. 
In the following, owing to the complexity of the equations, we will focus on the axial sector to explore the impact of the background plasma velocity field.

Schematically, the third-order equation governing the axial sector in the case of free-fall plasma reads
\begin{equation}
    A_1 u_{(4)}(r) + A_2 u_{(4)}'(r)+ A_3 u_{(4)}''(r)+ A_4 u_{(4)}'''(r)=0
\end{equation}
where the coefficients $A_i$ are given in Appendix~\ref{app:FFeq}. It is interesting to study the above equation in the limit of vanishing plasma frequency,
\begin{equation}\label{eq:axial_free_vanishing}
    4M\mathcal{D}_2 u_{(4)}+\left(2M\mathcal{D}_2 u_{(4)}-r(\mathcal{D}_2 u_{(4)})'  u_{(4)}'\right)=0 \, ,
\end{equation}
where the $\mathcal{D}_2$ operator was introduced in the previous section. From Eq.~\eqref{eq:finaleq}, we know that in this limit we must recover the vacuum Proca equation, $\mathcal{D}_2 u_{(4)}(r)=0$~\cite{Rosa:2011my}, which is indeed a solution to Eq.~\eqref{eq:axial_free_vanishing}. As we shall discuss in the next section, this is also the only solution compatible with the boundary conditions of the problem.

Finally, as the free-fall velocity vanishes at large distances, the field equations have the same asymptotic behavior at spatial infinity as in the case of an everywhere static plasma.

\subsubsection{Plasma density profile and plasma frequency}\label{sec:densityprofile}
The density of a plasma in free-fall can be obtained by solving the continuity equation~\eqref{eq:continuityeq} with four-velocity~\eqref{eq:FFvelocity}. This leads to a density profile of the form~\cite{Perlick:2015vta}  
\begin{equation}\label{eq:density}
    \rho(r)=\frac{\Dot{M}}{4\pi \sqrt{2M}}\frac{1}{r^{3/2}} \, ,
\end{equation}
where $\Dot{M}$ is a constant mass flux. This profile features an increasing density at the horizon, and vanishes at spatial infinity.
We can then express the plasma frequency as 
\begin{equation}
\label{eq:densityprofile}
  \omega_{\rm pl}^{\chi \, 2}=\omega_H^2\Big(\frac{2M}{r}\Big)^{3/2} \, ,
\end{equation}
where $\omega_H$ is the plasma frequency at the horizon, which in the following we will treat as a free parameter. 

To better compare perturbations in static and free-falling plasmas, we shall assume that the plasma frequency takes the form~\eqref{eq:densityprofile} also in the static case (where the continuity equation is satisfied for any time-independent $\rho$). 

\section{Quasibound spectrum of a Proca field in a plasma}
\label{sec:results}

\subsection{Numerical method and boundary conditions}
\label{sec:numerics}

In the following, we will solve the perturbation equations numerically using a direct integration shooting method~\cite{Ferrari:2007rc, Pani:2012bp, Rosa:2011my, Pani:2013pma}, wherein the system of radial equations is integrated from the horizon to infinity imposing suitable asymptotic conditions. This allows solving the eigenvalue problem and computing the complex eigenfrequencies of the modes $\omega=\omega_R+i\omega_I$. Given our convention for the time dependence of the eigenstates, $\sim e^{-i\omega t}$, stable modes corresponds to $\omega_I<0$. As already discussed, in the static case the system is composed of second-order differential equations, while in the free-fall case the field equations are of third differential order. Hence, for each equation, we will need two boundary conditions in the static case, and an extra one in the free-fall case.

\begin{figure*}[!t]
\centering
\includegraphics[width=0.48\textwidth]{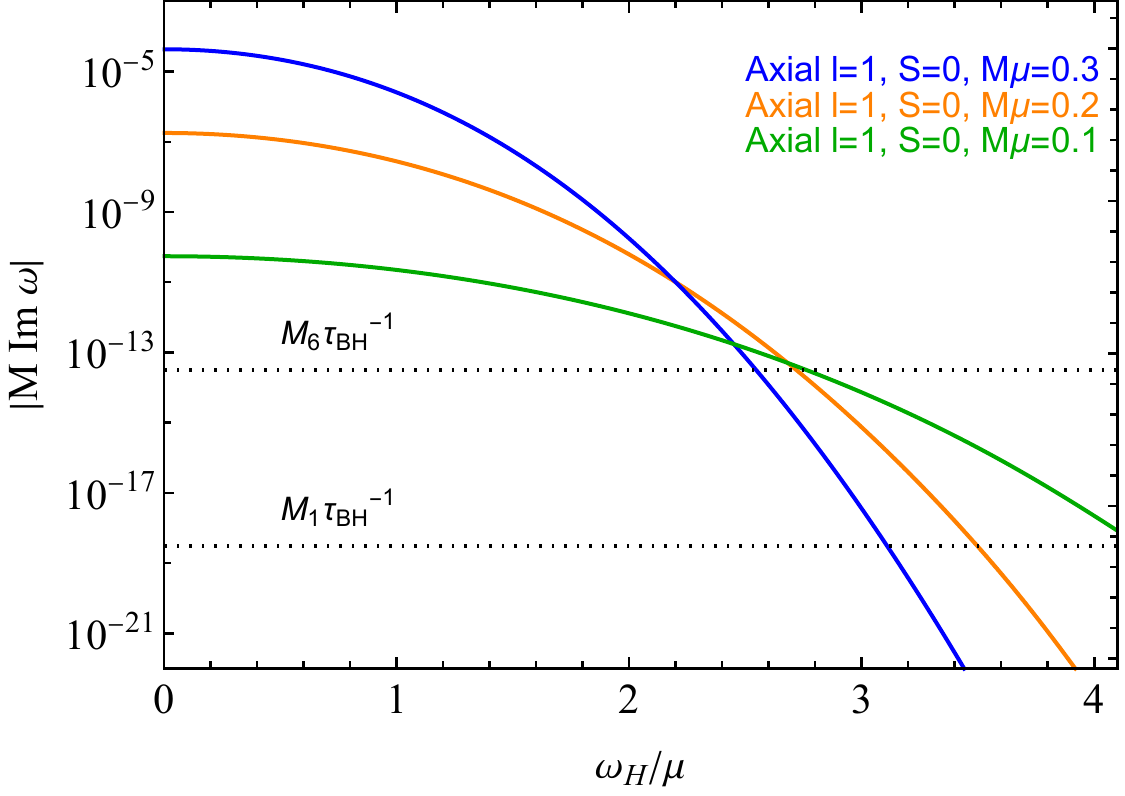}
\includegraphics[width=0.48\textwidth]{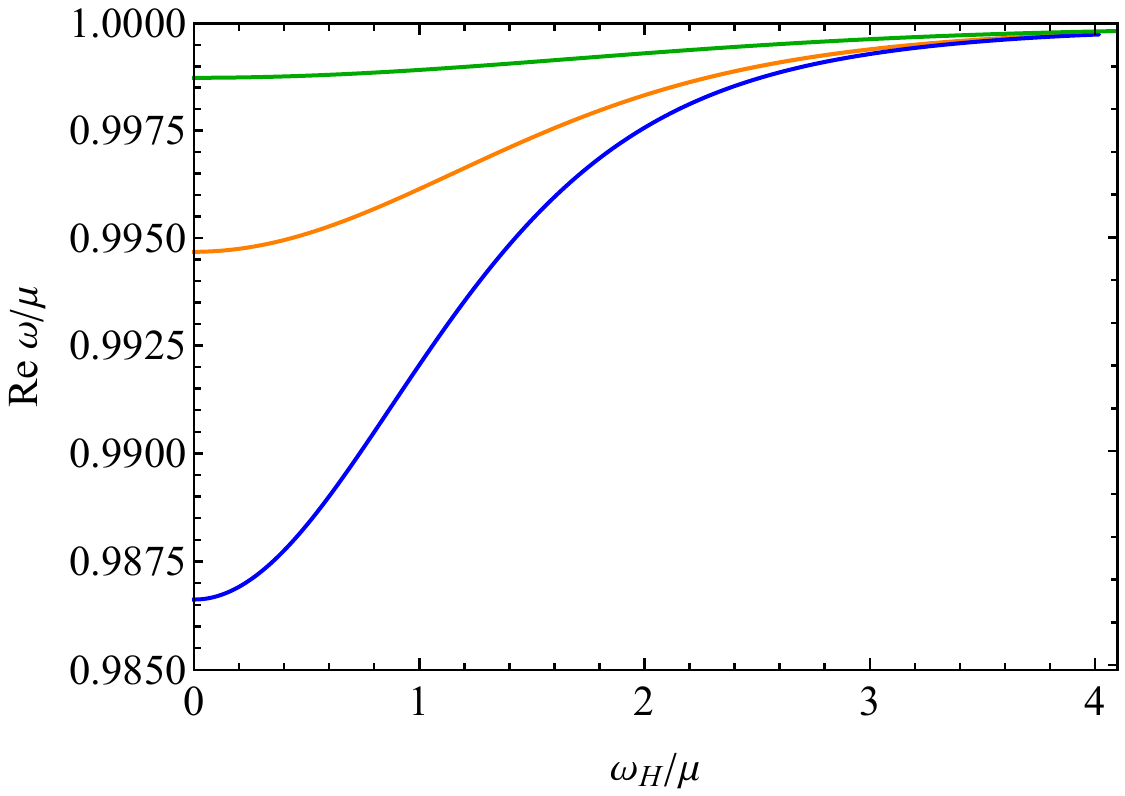}
\includegraphics[width=0.48\textwidth]{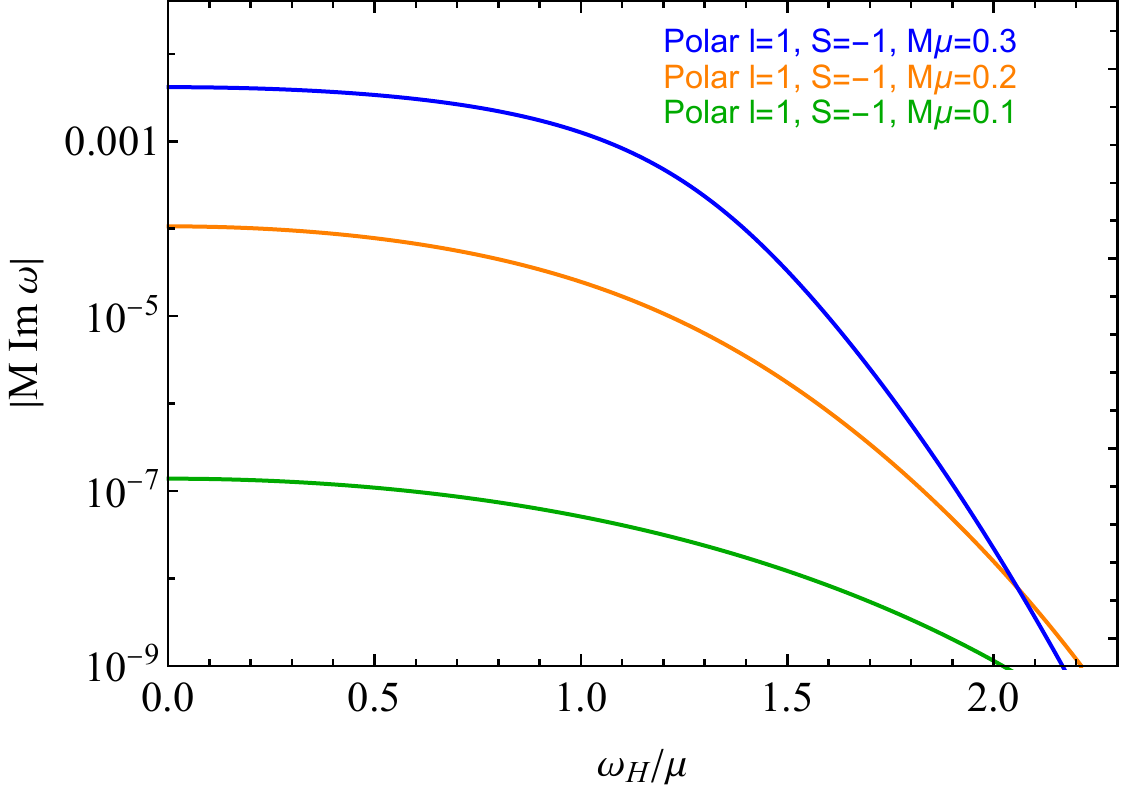}
\includegraphics[width=0.48\textwidth]{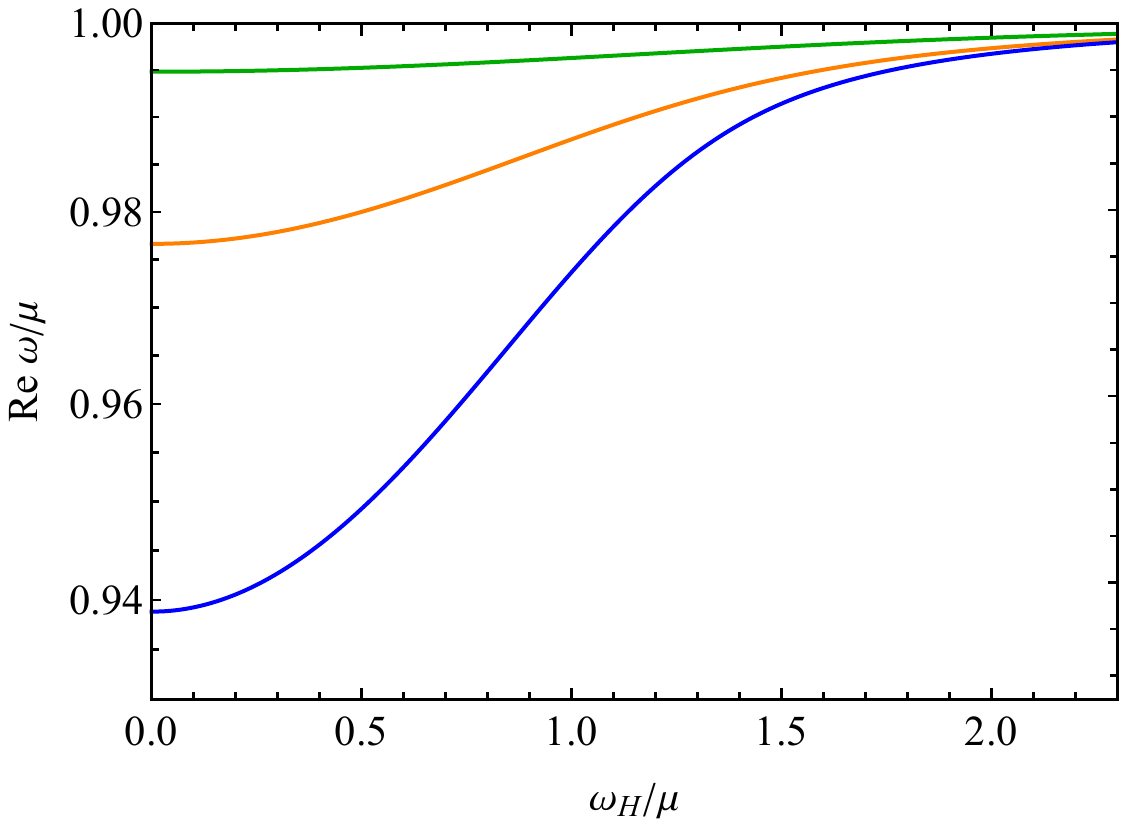}
\caption{(Upper panels) Imaginary (left) and real (right) part of the axial $l=1, S=0$ mode as a function of $\omega_H/\mu$ for different values of $\mu$ and for a static plasma. As the dark plasma frequency at the horizon increases, it effectively unbounds the modes: the binding energy $\omega_{\rm R}/\mu-1$ and the imaginary part $\omega_{\rm I}$ both go to zero, leading to larger timescales. 
(Lower panels) Same as in the upper panels but for the $l=1, S=-1$ dominant polar mode.}
\label{fig:ModeEvolution}
\end{figure*}

Let us start by analysing the standard static case.
Close to the horizon, we impose purely ingoing wave solutions, as the horizon behaves as a one-way membrane,
\begin{equation}
\label{eq:boundaryh}
    u_{(i)}\sim e^{-i\omega r_*}\sum_n b_{(i)\, n} (r-2M)^n,
\end{equation}
where the coefficients $b_{(i)\, n}$ can be computed in terms of the arbitrary coefficient $b_{(i)\, 0}$ by expanding the relevant equations near the horizon and solving them recursively. With the chosen density profile the plasma frequency vanishes at spatial infinity, see Eq.~\eqref{eq:densityprofile}, and therefore the leading-order solution can be generically written as a superposition 
\begin{equation}
u_{(i)}\sim B_{(i)}e^{-k_{\infty}r_* }+C_{(i)}e^{+k_{\infty}r_*},
\end{equation}
where $k_{\infty}=\sqrt{\mu^2-\omega^2}$. We are interested in finding solutions localised in the vicinity of the BH for $\omega<\mu$, i.e. solutions decaying at spatial infinity. Hence, we require $C_{(i)}=0$.

In the free-fall case, we can impose the same asymptotic conditions at infinity as in the static case. This is because the two configurations coincide in this limit, where the free-fall radial velocity vanishes.
On the other hand, by expanding the axial field equation at leading order at the horizon, we obtain 
\begin{equation}
\frac{d^2}{d r_*^2}u_{(4)}+b\frac{d}{d r_*}\Big(\omega^2 u_{(4)}+\frac{d^2}{d r_*^2}u_{(4)}\Big)+\omega^2 u_{(4)}=0 \, ,
\end{equation}
where $b$ is a constant. We can still impose that, at the leading order, the near-horizon solution is an ingoing wave, $u_{(4)}(r)\sim b_{(4)\, 0} e^{-i\omega r_*}$, where again $b_{(4)\, 0}$ is an arbitrary coefficient. However, in this case it is not possible to compute all the next-to-leading-order coefficients $b_{(i)\, n}$ solely in terms of $b_{(i)\, 0}$ by solving the field equations recursively as in the static-plasma case, since one coefficient is left unconstrained. To obtain the full solution near the horizon, we need a third, physically motivated, boundary condition. In particular, we must impose that the velocity perturbation of the fermions vanishes asymptotically at the horizon,
\begin{equation}
   \tilde{u}_\phi = \mathcal{O}(r-2M)\,. \label{3dBC}
\end{equation}
Indeed, in a free-fall plasma massive particles reach a background radial velocity equal to the speed of light at the horizon (as can be seen from Eq.~\eqref{eq:FFvelocity}), so any nonvanishing perturbation at the horizon would violate causality. This condition is automatically satisfied in the static case, where even in the vicinity of the BH horizon the plasma is static. 
As can be seen from Eq.~\eqref{eq:perturbedvelocity} in Appendix~\ref{app:Fieldeq}, the extra boundary condition~\eqref{3dBC} translates into a requirement on the component of the field orthogonal to the background four-velocity at the horizon. This third boundary condition provides the missing relation to obtain the asymptotic solution~\eqref{eq:boundaryh} in terms of a single arbitrary coefficient $b_{(i)\, 0}$.

\subsection{Spectrum of quasibound modes}

Figure~\ref{fig:ModeEvolution} shows the dependence of the axial (upper panels) and polar (lower panels) spectrum on the ratio $\omega_H/\mu$ between the dark plasma frequency at the horizon and the bare DP mass for different values of $\mu$ in a \textit{static} plasma\footnote{For the axial sector, we are able to solve for the dominant mode for larger values of $\omega_H/\mu$ and with higher precision, owing to the simplicity of the axial equation.}. The two sectors show the same behavior: as $\omega_H\rightarrow 0$, the plasma density goes to zero and we recover the results of Ref.~\cite{Rosa:2011my} describing the quasibound states of a noninteracting Proca field in a Schwarzschild spacetime. As $\omega_H$ becomes larger, the real part of the mode frequency increases rapidly toward the value of the bare mass, and therefore the binding energy of the modes, $\omega_{\rm R}/\mu-1$, vanishes. Meanwhile, the imaginary part decreases dramatically, leading to much larger timescales. Hence, the modes become more and more unbound. This behavior is due to the fact that as the effective mass increases at the horizon, the minimum in the effective potential flattens and therefore the formation of quasibound states is strongly hampered. The same phenomenology was recovered in a similar system in~\cite{Dima:2020rzg}.

When the effective mass is sufficiently large (which roughly occurs when $\sqrt{\omega_H^2+\mu^2} M\gg0.1$), we find an exponential decay of the imaginary part, as in the case of large bare mass~\cite{Zouros:1979iw}. In this case the modes become extremely long lived and of less astrophysical interest. Importantly, this would also affect the superradiant timescale over which the BH spin is dissipated, which is typically comparable to that of the quasibound states in the (stable) nonspinning limit~\cite{Brito:2015oca}. 

In the upper left panel of Fig.~\ref{fig:ModeEvolution} we show the values (dashed horizontal lines) of the (inverse) timescale $\tau_{\rm BH} \equiv 0.1 \, \tau_{\rm Salpeter}$, where $\tau_{\rm Salpeter} \simeq 4.5 \times 10^7$ years is the Salpeter timescale. This is the characteristic timescale of accretion of an astrophysical BH, and is the relevant one to compare with the superradiant timescale when deriving bounds from the superradiant instability~\cite{Brito:2015oca,Pani:2012vp,Pani:2012bp,Baryakhtar:2017ngi, Cardoso:2018tly, Ghosh:2021zuf} (see Sec.~\ref{sec:bound} for more details on experimental limits). In particular, if the superradiant timescale $ 1/|\omega_{\rm I}|$ is much longer than $\tau_{\rm BH}$ the superradiant instability is ineffective. 

As a rule of thumb, when $1/|\omega_{\rm I}|\gg\tau_{\rm BH}$, plasma effects are likely to completely invalidate superradiant bounds, since they destroy the quasibound states in the first place.
In Fig.~\ref{fig:ModeEvolution} we show two examples, for $M = 10^6 M_{\odot} \equiv M_6$ and $M = 10 M_{\odot} \equiv M_1$; in both cases if $\omega_H \gtrsim 2\, \text{--}\, 4 \, \mu$ the mode lifetimes are longer than the BH accretion timescale. The same happens for the polar mode (lower panels of  Fig.~\ref{fig:ModeEvolution}), although in this case it is numerically more challenging to push the modes towards large values of $\omega_H/\mu$, and therefore we do not show $\tau_{\rm BH}$ in that plot. Nevertheless, it is clear that also in the polar case the imaginary part of the modes becomes extremely small when $\omega_H \gtrsim 2\, \text{--}\, 4 \, \mu$, as in the axial case.

We conclude that the presence of a dark plasma, if sufficiently dense, can completely quench the quasibound spectrum of a DP and cause a dramatic increase of the mode lifetime. This would correspond to a severe weakening of the superradiant instability around a spinning BH.

\begin{figure}[h]
\centering
\includegraphics[width=0.49\textwidth]{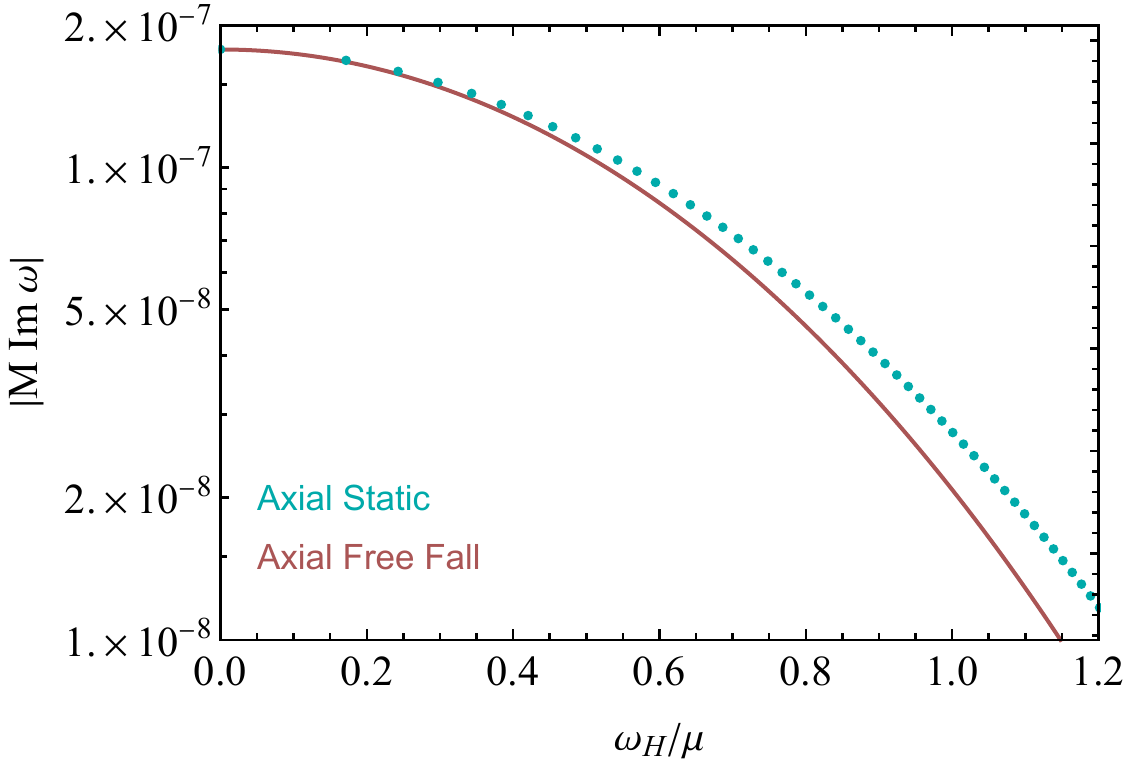}
\includegraphics[width=0.49\textwidth]{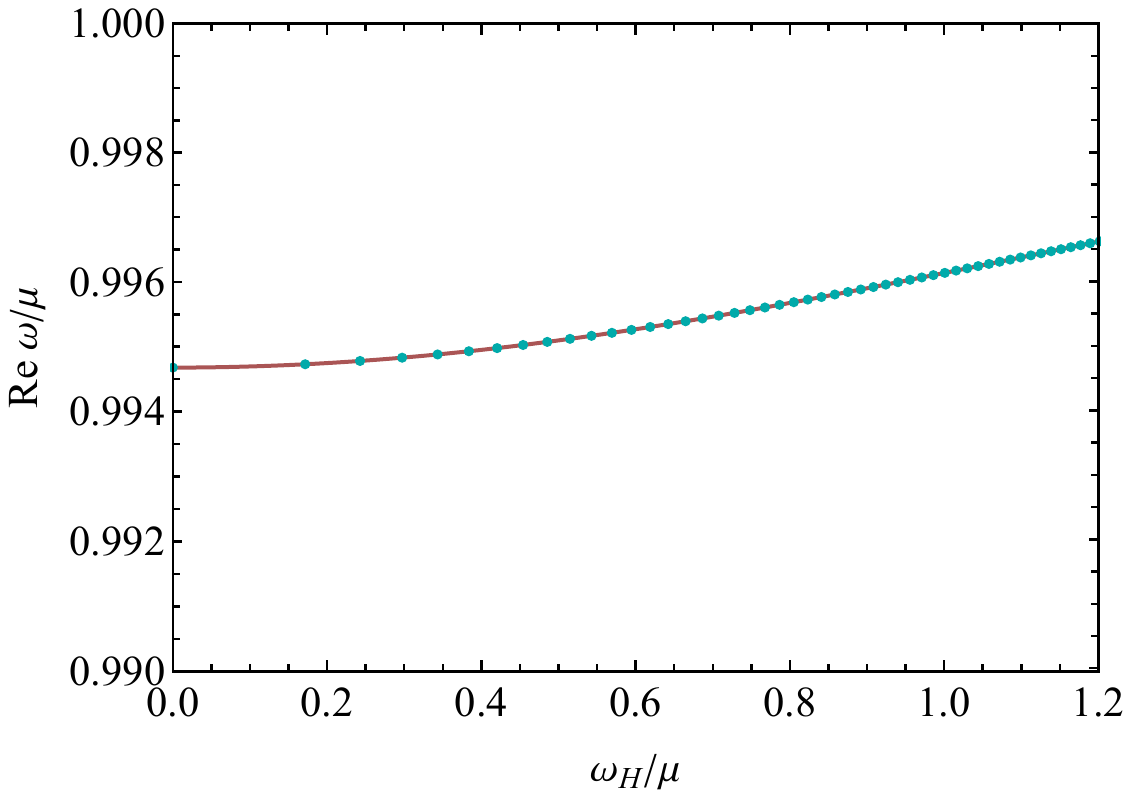}
\caption{Imaginary and real part of the axial $l=1, S=0$ mode as functions of the ratio $\omega_H/\mu$ with mass $M\mu=0.2$ for the static (dotted blue) and free-fall (red) case. For $M\omega_H=0$, i.e. in the absence of plasma, the two configurations coincide to the vacuum Proca axial mode. As $\omega_H/\mu$ increases, the real part of the mode is very similar for the two configurations, while the imaginary part of the mode decreases faster in the free-fall case. Overall, the two configurations exhibit a similar behavior.}
\label{fig:freefallaxial}
\end{figure}

Figure~\ref{fig:freefallaxial} shows a comparison between the dominant axial mode in the static and free-fall configurations for $M\mu=0.2$. For $\omega_H/\mu=0$, i.e. in the absence of plasma, the two modes coincide with the vacuum Proca axial mode. This is because, in the absence of plasma (when the dark plasma frequency goes to zero) the solution is trivially given by a vacuum Proca equation, see Eq.~\eqref{eq:finaleq}. Interestingly, as $\omega_H/\mu$ increases, the real part of the mode is essentially unaffected by the plasma motion, while the imaginary part decreases faster in the free-fall configuration. Overall, the free-fall configuration has the same phenomenology as the static one: as the dark plasma frequency increases, the binding energy goes to zero and the timescales are severely stretched. In the free-fall case, the drop in the imaginary part is slightly more severe. One can thus expect the superradiant instability to be even more severely weakened in this case. 

\begin{figure}[b!]
\centering
\includegraphics[width=0.49\textwidth]{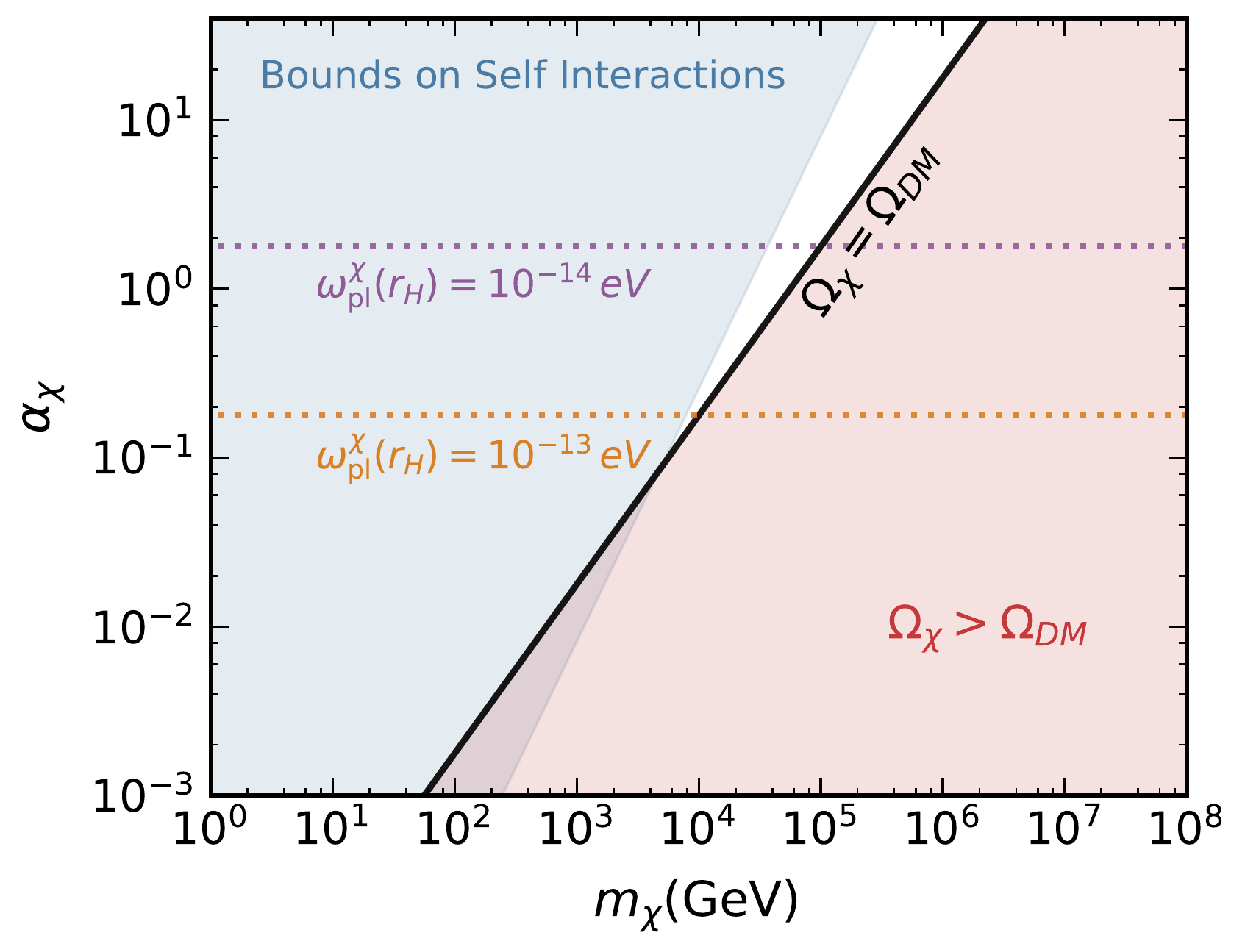}
\caption{Parameter space of interest for a simple secluded DM model. The blue shaded region is excluded because of the self-interactions bound from dwarf galaxies~\cite{Kaplinghat:2015aga}, while the solid black line indicates the parameters for which the correct relic abundance is obtained via freeze-out. The horizontal dotted lines indicates the value of $\alpha_\chi$ corresponding to two different plasma frequencies, assuming Bondi-Hoyle accretion around the BH horizon and the full DM density at ``infinite'' distance from it.}
\label{fig:DMSpace}
\end{figure}

\section{DM models}
\label{sec:DM}
In the previous sections we have shown that a dark plasma frequency comparable to, or larger than, the bare DP mass can greatly alter the quasibound states, possibly suppressing the DP superradiant growth. Here we sketch a simple DM model that can generate a sizable plasma frequency around a BH. The numerical conclusions of the previous sections are of course independent of the mechanism generating the fermionic relic abundance, and we could simply \textit{assume} the existence of a UV model generating the correct density at low redshift. Nevertheless, we find it useful to write down an illustrative model, which is not intended to be exhaustive of all possibilities. We hope our work motivates research into other alternatives in this direction. 

Let us consider a fermion $\chi$ with mass $m_\chi$ coupled to our DPs with fine structure constant $\alpha_\chi$. We consider $\chi$ to be a DM candidate and set its abundance via freeze-out. For simplicity, we study a ``secluded" scenario, so that couplings to the Standard Model thermal bath are not important for thermal freezeout~\cite{Pospelov:2007mp, Ackerman:2008kmp} and the relic abundance is set by the process $\bar{\chi} \chi \leftrightarrow V V$ (where $V$ schematically denotes the DP), which in the limit $\mu \ll m_\chi$ has a cross section 
\begin{equation}
    \langle \sigma v\rangle_{\bar{\chi} \chi \leftrightarrow V V} \simeq \frac{\pi \alpha_\chi^2}{m_\chi^2}. \label{eqsigma1}
\end{equation}
The DM relic abundance is obtained when the annihilation cross section is of the order
\begin{equation} \label{eqsigma2}
\langle \sigma v\rangle_{\bar{\chi} \chi \leftrightarrow V V} \simeq \frac{1}{10^9 \rm GeV^2},
\end{equation}
where we took $T_{\rm fo} \simeq m_\chi/10$ for the freezeout temperature and assumed typical values for the relativistic degrees of freedom in the early universe~\cite{Lin:2019uvt}. 

The dark fermions in this model will exhibit self-interactions due to the exchange of a DP. Self-interactions are especially enhanced in the limit of small DP mass, as the cross section presents a forward scattering enhancement. In fact, in the Born limit the transfer cross section for a DM particle of velocity $v_{\chi}$ reads
\begin{equation}
    \sigma_{\rm self} \simeq \frac{8 \pi \alpha_\chi^2}{m_\chi^2 v_\chi^4} \ln\Big(\frac{m_{\chi}^2 v_\chi^2}{\mu^2}\Big).
\end{equation}
Limits on DM self-interactions (SIDM) from observations of galaxy clusters, galaxies, and dwarf-galaxy halos~\cite{Kaplinghat:2015aga, Kaplinghat:2015aga, Andrade:2020lqq, Ackerman:2008kmp} restrict this cross section to be roughly $\sigma_{\rm self} / m_\chi \lessapprox 0.1 - 100 \, \rm cm^2/g$. 

Comparing the couplings needed to set the relic abundance, one can see easily that the SIDM constraint excludes thermal freezeout for DM masses smaller than $m_\chi \sim \rm TeV$. This is shown in Fig.~\ref{fig:DMSpace}. The blue shaded area indicates the region excluded by measurements of dwarf galaxies~\cite{Kaplinghat:2015aga}, which limit the cross section to be $\sigma_{\rm self} / m_\chi \lesssim 10-100 \, \rm cm^2/g$ for DM velocities $v_{\chi} \sim 10^{-4}$ (here we also fixed $\mu = 10^{-14} \rm eV$, but the dependence on the DP mass is very weak). The solid black line indicates the parameters for which the correct relic abundance is obtained via freeze-out. The horizontal dotted lines indicates the value of $\alpha_\chi$ corresponding to two different plasma frequencies, defined as usual as
\begin{multline}
\label{Eq:PlasmaDark}
    \omega_{\rm pl}^\chi = \Big(\frac{4\pi \rho_\chi \alpha_\chi}{m_\chi^2}\Big)^{1/2} \simeq \\ \simeq 1.8 \times 10^{-13} \textrm eV \Big(\frac{\rho_\chi}{10^5 \rm \, GeV/cm^3}\Big)^{1/2}\Big(\frac{0.1}{\alpha_\chi}\Big)^{1/2},
\end{multline}
where in the last step we used Eqs.~\eqref{eqsigma1} and \eqref{eqsigma2} to relate $m_\chi$ to $\alpha_\chi$\footnote{Notice that the second line of Eq.~\eqref{Eq:PlasmaDark} is valid \textit{only} if the freeze-out happens via Eq.~\eqref{eqsigma1}. If other interactions, independent of $\alpha_\chi$, set the relic abundance, then the full expression of the plasma frequency (with no a priori relation between $\alpha_\chi$ and $m_\chi$) should be used.}. The normalization of $\rho_\chi$ is estimated by assuming Bondi-Hoyle accretion, which gives the following density close to the BH horizon
\begin{equation}
    \rho_\chi(r_{\rm H}) \simeq 2.4 \times 10^5 \rho_\chi^{\infty} \Big(\frac{0.01}{v_{\rm rel}}\Big)^3, 
\end{equation}
where $v_{\rm rel}$ is the relative BH-DM velocity far from the BH and $\rho_\chi^{\infty}\approx 0.4 {\rm GeV/cm}^3$ is the DM ambient density far away from the horizon.

From Fig.~\ref{fig:DMSpace} we see that, within this minimal model, dark fermions can naturally dress the DP with a plasma mass, but the available parameter space is small and confined to generate a plasma frequency of roughly $\sim 10^{-13}\,{\rm eV}$, if we require the dark fermion to be in the perturbative regime ($\alpha_\chi\lesssim 1$), have the correct relic abundance, and avoid SIDM bounds. However, as already mentioned, this simple, minimal model is easily extendable to widen the allowed parameter space. This can be achieve by: i)~relaxing self-interaction bounds, or/and ii) producing the DM relic abundance through a different interaction.  
One possibility to relax SIDM bounds is to have a large splitting for the Dirac fermion, which kinematically suppresses self-interactions~\cite{Schutz:2014nka, Vogelsberger:2018bok}. 
A simple extension to evade the relic abundance requirement is described in Ref.~\cite{Ackerman:2008kmp}, where the DM fermion is also charged under $SU(2)_{\rm L}$. In this case, the relic abundance can be set at early times by freeze-out via the weak interaction. At late times, the weak cross section remains small, while the long-range cross section mediated by the DP comes to dominate as the DM cools and slows, reducing to the vanilla model described here. In this way, it can be possible to extend to lower DM masses, increasing the allowed plasma frequency.

Finally, the DM density around the BH may be much larger than the value set by Bondi-Hoyle accretion, possibly leading to even larger plasma frequencies. This would be the case, for example, if DM spikes are present around the BH~\cite{Ullio:2001fb, Gondolo:1999ef}.

\section{Impact of DP-DM coupling on current DP bounds from BH superradiance}
\label{sec:bound}

\begin{figure}[b!]
\centering
\includegraphics[width=0.49\textwidth]{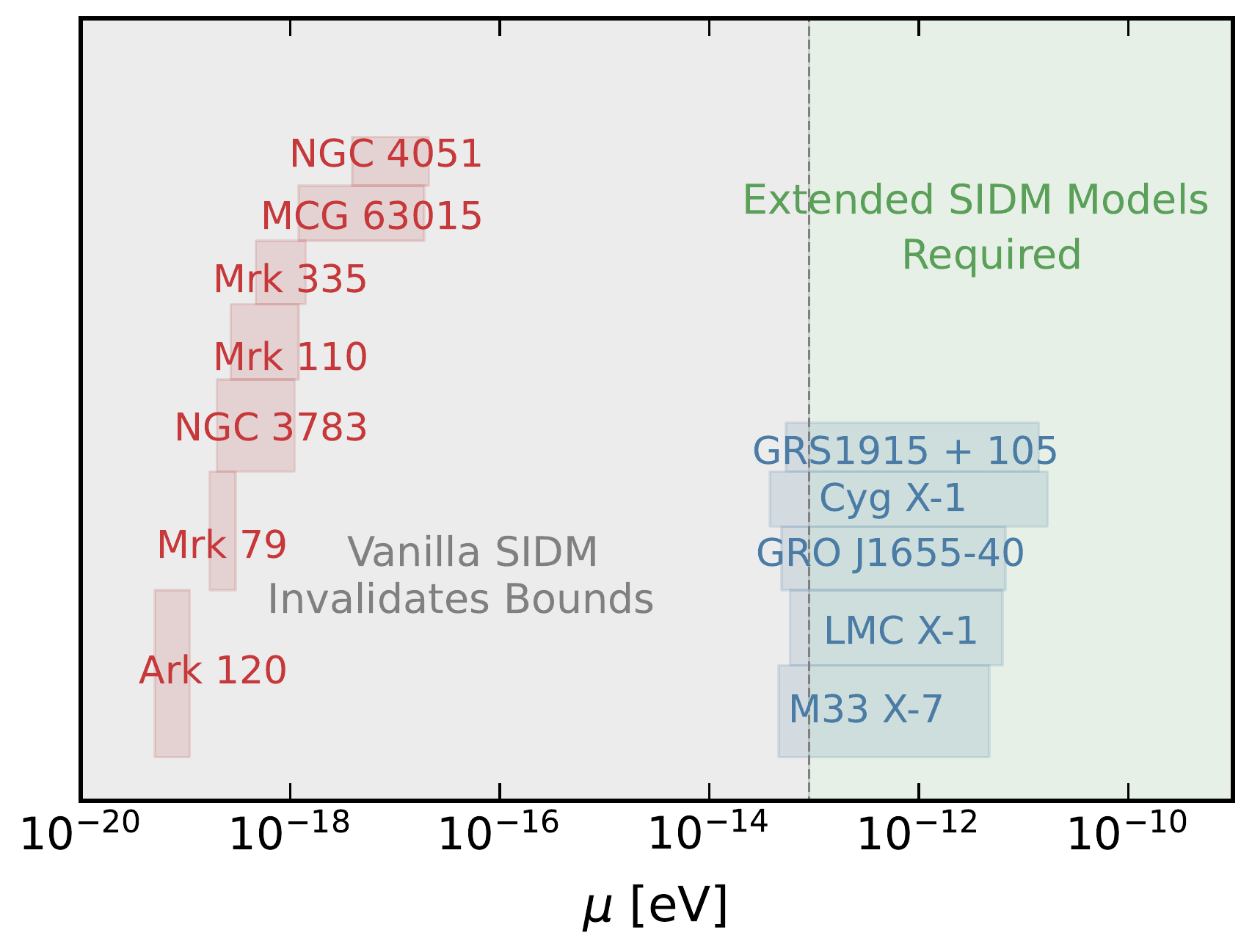}
\caption{Constraints on DP masses derived from highly-spinning supermassive (red bands) and stellar-mass (blue bands) BHs as computed in Ref.~\cite{Baryakhtar:2017ngi}. The gray region indicates the masses for which the simple SIDM model of Sec.~\ref{sec:DM} can introduce strong plasma effects which invalidate the bounds. The green region indicates masses for which extensions of the simplest model are required in order to generate a sizable enough $\omega_{\rm pl}^\chi$ without overproducing DM and without violating SIDM bounds.}
\label{fig:BHBounds}
\end{figure}

Probing ultralight DPs is an extremely challenging task, as lab-based experiments become impractical for very long wavelengths~\cite{Caputo:2021eaa}. Most of the constraints on DP masses smaller than $\mu\sim 10^{-10}$ eV then come from astrophysics and cosmology, typically assuming a kinetic mixing between the DP and the ordinary photon~\cite{Caputo:2020bdy, Caputo:2020rnx, Bondarenko:2020moh}. In this regard, superradiance represents a unique possibility to probe very light DPs even without assuming a kinetic mixing to the SM photon. 

In the past years, many authors have put tight constraints on DPs using the BH mass-spin distribution (see, e.g., Refs.~\cite{Arvanitaki:2010sy,Pani:2012vp,Pani:2012bp,Baryakhtar:2017ngi,Brito:2017wnc,Brito:2017zvb,Cardoso:2018tly,Ng:2019jsx,Fernandez:2019qbj,Ghosh:2021zuf} and \cite{Brito:2015oca} for a recent summary of the constraints). 
The basic physics in all these cases is the same: if a DP with the correct mass is present in the spectrum of the theory, a superradiant growth can be triggered on short-enough timescales and extract angular momentum and energy from astrophysical BHs, making the measured spins inconsistent with the DP itself, and also leading to peculiar gravitational-wave signatures~\cite{Brito:2015oca}. In particular, current stellar-origin BH mass-spin measurements in X-ray binaries\footnote{Note that constraints from X-ray binaries might be affected by systematic errors in the determination of the BH spin, see discussion in Ref.~\cite{Belczynski:2021agb}.} exclude DP masses between $\mu \sim 5 \times 10^{-14} \text{--} \, 2 \times 10^{-11} \,{\rm eV}$, while supermassive BH spin measurements exclude lighter masses $\mu \sim \,6 \times 10^{-20} \text{--} \, 2 \times 10^{-17} {\rm eV}$~\cite{Pani:2012vp,Pani:2012bp,Baryakhtar:2017ngi,Cardoso:2018tly,Ghosh:2021zuf}. These bounds are reported in Fig.~\ref{fig:BHBounds}, where the red bands refer to highly-spinning supermassive BHs, while the blue ones to stellar-mass BHs (see Ref.~\cite{Baryakhtar:2017ngi} for a similar plot). 

In order for these bounds to apply, the superradiant rate should be fast enough to grow a maximally-filled cloud within the relevant BH timescale, $\tau_{\rm BH}$. Our present analysis shows that, in motivated DM models, it may be very easy to make the superradiant timescale larger than $\tau_{\rm BH}$. In particular, within the simple class of DM models described in the previous section, we have shown that it is possible to obtain plasma frequencies of the order $\omega_{\rm pl}^\chi \sim 10^{-13}$ eV, at least. This means that DPs a factor of few lighter than this are easily rescued thanks to the plasma effects discussed in Sec.~\ref{sec:results} (see Fig.~\ref{fig:ModeEvolution}, \textit{e.g.}, upper left panel). This corresponds to the gray region in Fig.~\ref{fig:BHBounds}, where the vertical line is drawn using the reference value in Eq.~\eqref{Eq:PlasmaDark}, divided by a factor of 2 to account for the onset of the effect observed in Fig.~\ref{fig:ModeEvolution}. 

As outlined in Sec.~\ref{sec:DM}, it seems also likely that simple extensions of this model could rescue even larger DP masses, corresponding to the green region in Fig.~\ref{fig:BHBounds}. 

Finally, we stress that our argument can invalidate superradiance bounds while leaving unchanged cosmological ones in the same mass range. The latter typically rely on resonant effects on large scales~\cite{Caputo:2020bdy}, where the dark plasma density is several orders of magnitude smaller than around a BH. 

\section{Conclusion and extensions}
\label{sec:discussion}
In this work, we have studied for the first time the quasibound states of a DP field in the presence of a (dark) plasma. The latter dresses the DP with an additional density-dependent mass, which can significantly alter the quasibound spectrum. In particular, we showed that if the generated plasma frequency is $\gtrsim 2\, \text{--}\, 4$ times the DP bare mass, then the state lifetime increases dramatically and (extrapolating to spinning BHs) the superradiant instability is effectively quenched, similarly to the case of bosons with a large bare mass ($M\mu\gg1$~\cite{Zouros:1979iw}) which is of limited astrophysical interest. This has important implications for observational bounds on DPs. We outlined a simple, motivated particle physics model that naturally provides a sizable plasma frequency, possibly bearing a way out from current superradiance bounds from BH mass-spin measurements.

Interestingly, our analysis also shows that if current or future detectors will discover gravitational-wave signatures of a new light vector particle through BH superradiance~\cite{Pani:2012vp,Pani:2012bp,Baryakhtar:2017ngi,Cardoso:2018tly,Siemonsen:2019ebd,Tsukada:2020lgt} this could be used to set relevant constraints on the DP scenario in various motivated DM models.

\begin{acknowledgments}

P.P. acknowledges financial support provided under the European Union’s H2020 ERC, Starting Grant agreement no. DarkGRA–757480. We also acknowledge support under the MIUR PRIN and FARE programmes (GW-NEXT, CUP: B84I20000100001), and from the Amaldi Research Center funded by the MIUR program ``Dipartimento di Eccellenza'' (CUP: B81I18001170001). AC is supported by the Foreign Postdoctoral Fellowship Program of the Israel Academy of Sciences and Humanities and also acknowledges support from the Israel Science Foundation (Grant 1302/19) and the European Research Council (ERC) under the EU Horizon 2020 Programme (ERC-CoG-2015-Proposal n. 682676 LDMThExp).  
This work is partially supported by the PRIN Grant 2020KR4KN2 ``String Theory as a bridge between Gauge Theories and Quantum Gravity''.

\end{acknowledgments}

\bibliographystyle{utphys}
\bibliography{Ref}

\onecolumngrid
\appendix


\setcounter{equation}{0}
\setcounter{figure}{0}
\setcounter{table}{0}
\setcounter{page}{1}
\makeatletter
\renewcommand{\theequation}{S\arabic{equation}}
\renewcommand{\thefigure}{S\arabic{figure}}
\renewcommand{\thepage}{S\arabic{page}}

\begin{center}
\textbf{\large Appendix\\[0.5ex]
}
\end{center}

In these Appendices we provide the technical details of our analysis. In particular, we write down all the field equations for both the static and the free-fall plasma. 

\bigskip

\twocolumngrid

\section{The master differential equation}\label{app:Fieldeq}
In the following, we will rearrange the system of Eqs.~\eqref{eq:perturbedprocaeq}-\eqref{eq:perturbednorm} into a single master equation for the DP field. To do so, let us define the following projection operator 
\begin{equation}
    h^{\alpha\beta}=g^{\alpha\beta}+ u^\alpha u^\beta \, ,
\end{equation}
which projects vectors onto hypersurfaces whose normal vector is the fermionic four velocity. Then, we can decompose the derivative of the four-velocity as~\cite{Ellis:1971pg} 
\begin{equation}
\label{eq:velocitydecomposition}
    \nabla_\mu u^\nu=\theta_{\ \ \mu}^{ \nu}+\omega_{\ \ \mu}^{ \nu}-u_\mu u^\rho \nabla_\rho u^\nu \, ,
\end{equation}
where the tensors $\theta^{\alpha\beta}$ and $\omega^{\alpha\beta}$ are the deformation and vorticity tensors, defined as the symmetric and antisymmetric part of the tensor $v^{\mu\nu}=h^{\mu \alpha }h^{\nu \beta }u_{\alpha;\beta}$: 
\begin{eqnarray}
 \theta_{\mu\nu} &=& \frac{1}{2}(v_{\mu\nu}+v_{\nu\mu})\,, \label{deform}\\
 \omega_{\mu\nu} &=& \frac{1}{2}(v_{\mu\nu}-v_{\nu\mu})\,. \label{vort}
\end{eqnarray}
Finally, we can decompose the background field strength in to an electric component $E^\nu \equiv F^{\nu\mu}u_\mu$ and a magnetic one $B_{\mu\nu} \equiv h_{\mu}{}^{\alpha}h_{\nu}{}^{\beta}F_{\alpha \beta}$, which leads to a definition of the Larmor tensor for the DP: ${\omega_{\rm L}}^{\mu\nu} = -\frac{q_\chi}{m_\chi} B^{\mu\nu}$. 

To reassemble the system of equations into a single one, we can express the perturbed four-velocity in terms of the DP field by projecting Eq.~\eqref{eq:perturbedprocaeq}, 
\begin{equation}
\label{eq:perturbedvelocity}
 \Tilde{u}^\rho=\frac{h^\rho_{\ \nu}}{q_\chi n}(\nabla_\mu \tilde{F}^{\nu\mu}+\mu^2 \tilde{V}^\nu) \, ,
\end{equation}
and insert it into Eq.~\eqref{eq:perturbedmomentumeq} to obtain
\begin{multline}
    \label{eq:intermediateeq}
    h^{\mu}_{\ \ \alpha}(\nabla_\sigma \tilde{F}^{\alpha \sigma }+\mu^2 \tilde{V}^\alpha)\nabla_\mu u^\nu+u^{\mu}\nabla_\mu h^{\nu}_{\ \ \alpha} (\nabla_\sigma \tilde{F}^{\alpha \sigma }+\mu^2 \tilde{V}^\alpha)\\-\frac{1}{n}u^{\mu}\partial_\mu n \  h^{\nu}_{\ \ \alpha}(\nabla_\sigma \tilde{F}^{\alpha \sigma }+\mu^2 \tilde{V}^\alpha)=  \omega_{\rm pl}^{\chi \, 2}\tilde{F}^{\nu\mu}u_\mu\\+\frac{ q_\chi}{m_\chi}F^{\nu\mu}h_{\mu\alpha}(\nabla_\sigma \tilde{F}^{\alpha \sigma }+\mu^2 \tilde{V}^\alpha)
\end{multline}
where we defined a plasma frequency for the oscillations induced by the DP propagation in the plasma, $ \omega_{\rm pl}^{\chi \, 2}=q_\chi^2 n/m_\chi$.

We still want to rearrange this equation into a more convenient form, to obtain Eq.~\eqref{eq:finaleq}. In the following, we provide a step-by-step calculation, focusing individually on each term of Eq.~\eqref{eq:intermediateeq}. Let us start by handling the first term: we decompose the first derivative of the four velocity in the standard way, via Eq.~\eqref{eq:velocitydecomposition},
\begin{multline}
  (\nabla_\sigma \tilde{F}^{\alpha \sigma }+\mu^2 \tilde{V}^\alpha)(g^\mu_{\ \ \alpha}+u^\mu u_\alpha)\nabla_\mu u^\nu= \\(\nabla_\sigma \tilde{F}^{\alpha \sigma }+\mu^2 \tilde{V}^\alpha)(g^\mu_{\ \ \alpha}+u^\mu u_\alpha)(\theta_{\ \ \mu}^{ \nu}+\omega_{\ \ \mu}^{ \nu}-u_\mu u^\rho \nabla_\rho u^\nu) \, .
\end{multline}
Next, we use the following identities: $u^\mu\theta_\mu^{\ \ \nu}=u^\mu\omega_\mu^{\ \ \nu}=0 $
and $(g^\mu_{\ \ \alpha}+u^\mu u_\alpha)(-u_\mu u^\rho \nabla_\rho u^\nu)=0$ (where the latter follows from $u^\mu u_\mu=-1$) so that the first term simply becomes
\begin{equation}
\label{eq:firsttermintermediate}
    (\theta_{\ \ \alpha}^{\nu}+\omega_{\ \ \alpha}^{\nu})(\nabla_\sigma \tilde{F}^{\alpha \sigma }+\mu^2 \tilde{V}^\alpha) \, .
\end{equation}
Now let us manipulate the second term of (\ref{eq:intermediateeq}).
We have
\begin{align}
   & u^\mu \nabla_\mu h^{\nu}_{\ \ \alpha}(\nabla_\sigma \tilde{F}^{\alpha \sigma }+\mu^2 \tilde{V}^\alpha)=u^\mu h^{\nu}_{\ \ \alpha}\nabla_\mu (\nabla_\sigma \tilde{F}^{\alpha \sigma }+\mu^2 \tilde{V}^\alpha) \nonumber\\
   &+(\nabla_\sigma \tilde{F}^{\alpha \sigma }+\mu^2 \tilde{V}^\alpha) u^\mu (u^\alpha \nabla_\mu u^\nu+u^\nu \nabla_\mu u_\alpha)\,.
\end{align}
As for the third term, we shall use the continuity equation~\eqref{eq:continuityeq} to get
\begin{multline}
\label{eq:thirdtermintermediate}
    -\frac{1}{n}u^{\mu}\partial_\mu n \  h^{\nu}_{\ \ \alpha}(\nabla_\sigma \tilde{F}^{\alpha \sigma }+\mu^2 \tilde{V}^\alpha)\\
    =\frac{1}{n}n\nabla_\mu u^\mu \  h^{\nu}_{\ \ \alpha}(\nabla_\sigma \tilde{F}^{\alpha \sigma }+\mu^2 \tilde{V}^\alpha)\\
    =\theta h^{\nu}_{\ \ \alpha}(\nabla_\sigma \tilde{F}^{\alpha \sigma }+\mu^2 \tilde{V}^\alpha) \, ,
\end{multline}
where $\theta=\theta_\mu ^\mu$ is the trace of the deformation tensor. As for the fourth and fifth terms on the right side of Eq.~\eqref{eq:intermediateeq}, we just leave them in their original form.

Let us now apply the operator $h^\xi_{\ \ \nu}$ on every term. From (\ref{eq:firsttermintermediate}) it is easy to see that the first term becomes simply $ (\theta_{\ \ \alpha}^{\xi}+\omega_{\ \ \alpha}^{\xi})(\nabla_\sigma \tilde{F}^{\alpha \sigma }+\mu^2 \tilde{A}^\alpha)$,  as the deformation and vorticity are orthogonal to the four-velocity.
As for the second term, we can use $h^\xi_{\ \ \nu}u ^\nu=0$ and $h^\xi_{\ \ \nu}h^\nu_{\ \ \alpha}=h^{\xi}_{\ \ \alpha}$. We thus have
\begin{multline}
 u^\mu h^{\xi}_{\ \ \alpha} \nabla_\mu (\nabla_\sigma \tilde{F}^{\alpha \sigma }+\mu^2 \tilde{V}^\alpha)\\
 +(\nabla_\sigma \tilde{F}^{\alpha \sigma }+\mu^2 \tilde{V}^\alpha) u^\mu u_\alpha h^{\xi}_{\ \ \nu}\nabla_\mu u^\nu=\\ 
 u^\mu h^{\xi}_{\ \ \alpha} \nabla_\mu (\nabla_\sigma \tilde{F}^{\alpha \sigma }+\mu^2 \tilde{V}^\alpha)+(\nabla_\sigma \tilde{F}^{\alpha \sigma }+\mu^2 \tilde{V}^\alpha) u_\alpha u^\mu \nabla_\mu u^\xi \, ,
\end{multline}
where we used $u^\nu \nabla_\mu u_\nu=0$. By using the momentum equation on the second term $ u^\mu \nabla_\mu u^\xi$ we obtain
\begin{equation}
    h^{\xi}_{\ \ \alpha} u^\mu \nabla_\mu (\nabla_\sigma \tilde{F}^{\alpha \sigma }+\mu^2 \tilde{V}^\alpha)+\frac{q_\chi}{m_\chi}u_\alpha F^{\xi \beta}u_\beta (\nabla_\sigma \tilde{F}^{\alpha \sigma }+\mu^2 \tilde{V}^\alpha)\,.
\end{equation}
 As for the third term, from (\ref{eq:thirdtermintermediate}) it is simply
\begin{equation}
    \theta h^{\xi}_{\ \ \alpha}(\nabla_\sigma \tilde{F}^{\alpha \sigma }+\mu^2 \tilde{V}^\alpha)\,,
\end{equation}
where we used $h^\xi_{\ \ \nu}h^\nu_{\ \ \alpha}=h^\xi_{\ \ \alpha}$.
The fourth term simply becomes
\begin{multline}
    h^\xi_{\ \ \nu} \omega_{\rm pl}^{\chi \, 2}\tilde{F}^{\nu\mu}u_\mu \\
    = \omega_{\rm pl}^{\chi \, 2}\tilde{F}^{\xi\mu}u_\mu+ \omega_{\rm pl}^{\chi \, 2} u^\xi u_\nu\tilde{F}^{\nu\mu}u_\mu= \omega_{\rm pl}^{\chi \, 2}\tilde{F}^{\xi\mu}u_\mu \, ,
\end{multline}
where we used the anti-symmetric nature of the field strength. As for the fifth and last term of Eq.~\ref{eq:intermediateeq}, upon projection it gives
\begin{equation}
    \frac{q_\chi}{m_\chi}h^\xi_{\ \ \nu}F^{\nu\mu}h_{\mu\alpha}(\nabla_\sigma \tilde{F}^{\alpha \sigma }+\mu^2 \tilde{V}^\alpha)\,.
\end{equation}
Since by the definition $B_{ab}=h_a^{\ \ c}h_b^{\ \ d}F_{c d}$ and $\omega_{L \ ab}=-\frac{ e}{m}B_{ab}$, 
we can rewrite this term as
\begin{equation}
    -\omega^{\ \xi}_{L \ \  \alpha}(\nabla_\sigma \tilde{F}^{\alpha \sigma }+\mu^2 \tilde{V}^\alpha)\,.
\end{equation}
Assembling all terms together leads to the final Eq.~\eqref{eq:finaleq}, describing the propagation of a massive spin-1 particle in a cold, collisionless plasmic medium in curved spacetime.

\section{Field equations in the multipolar expansion}
\subsection{Static case}\label{app:Staticeqs}
We assume an unmagnetised background plasma, ${\omega_{\rm L}}^{\mu\nu}= 0$. The four velocity of a static plasma 
is $u^\alpha=(f^{-1/2}, \vec{0})$, satisfying the normalisation condition $u_{\mu}u^{\mu} = -1$. From Eq.~\eqref{eq:momentumeq}, the electric field has then only one nonvanishing radial component $E^\alpha=(0, m_\chi/q_\chi \, \Gamma^r_{00}(u^0)^2,0,0)$, where $\Gamma^\mu_{\alpha\beta}$ are the Christoffel's symbols.
In this case it can be seen that the vorticity and deformation tensors are both zero, $\omega^{\alpha\beta}=\theta^{\alpha\beta}=0$. 
By performing the multipolar expantion and working in the frequency domain we obtain the following set of equations:
\begin{align}
    \nonumber -\omega (f(l(l+1)+r^2\mu^2)-r^2\omega^2+f r^2 \omega_{\rm pl}^{\chi \, 2}) u_{(2)}&\\\nonumber +i f r (\omega^2 - f  \omega_{\rm pl}^{\chi \, 2}) u_{(1)}&\\+f r (-i r (\omega^2 - f  \omega_{\rm pl}^{\chi \, 2}) u_{(1)}' +f \omega u_{(3)}')&=0 \, ,\\
   \nonumber-l(l+1)r (\omega^2 - f  \omega_{\rm pl}^{\chi \, 2})u_{(1)}- i f l(l+1)\omega  u_{(2)}&\\\nonumber + i r^2 \omega (f \mu^2 -\omega^2 +f  \omega_{\rm pl}^{\chi \, 2}) u_{(3)}&\\ + i f \omega (l(l+1) r u_{(2)}' -2M u_{(3)}'- f r^2 u_{(3)}''))&=0 \, ,\\
    (f(l+l^2+r^2\mu^2)-r^2\omega^2+f r^2  \omega_{\rm pl}^{\chi \, 2})u_{(4)}&\nonumber\\-f(2M u_{(4)}'+f r^2 u_{(4)}'')&=0 \, .
\end{align}
where $u_{(i)}'=\partial_r u_{(i)}$, we have suppressed the $l$ superscript, and the radial dependence of $\omega^\chi_{\rm pl}=\omega^\chi_{\rm pl}(r)$. Now we have three equations for the four wavefunctions $u_{(1)}, u_{(2)}, u_{(3)}, u_{(4)}$. We can close the system with the Lorenz condition
\begin{equation}
\label{eq:lorenzcond}
    -ir^2\omega u_{(1)}-r f(u_{(2)}-u_{(3)}+r u_{(2)}')=0 \, .
\end{equation}
By solving the Lorenz equation for $u_1$ and plugging it in to the polar field equations we obtain Eqs.~\eqref{eq:D2eq},~\eqref{eq:D3eq}.

\subsection{Free-fall case}\label{app:FFeq}
In the free-fall configuration, plasma particles follow geodesics, i.e. the background DP field is set to $F^{\mu\nu}=0$, $E^\alpha=\omega_L^{\alpha\beta}=0$. 

A free-fall plasma does not have vorticity, $\omega^{\alpha\beta}=0$, but has a nonvanishing deformation. The nonzero components are 
\begin{align}
    &\theta_0^{\ 0}=\frac{\sqrt{2}(M/r)^{3/2}}{2M-r}, \quad
    \theta_r^{\ r}=\frac{\sqrt{M/r}}{\sqrt{2}(r-2M)}, \quad
    \theta_\theta^{\ \theta}=-\frac{\sqrt{2M}}{r^{3/2}}, \nonumber \\&
    \theta_\phi^{\ \phi}=-\frac{\sqrt{2M}}{r^{3/2}}, \quad
    \theta_0^{\ r}=\frac{M}{r^2}, \quad
    \theta_r^{\ 0}=-\frac{M}{(r-2M)^2} \, .
\end{align}
The trace of this tensor is also different from zero and therefore there are extra terms in Eq.~\eqref{eq:finaleq} with respect to the static case. Working again in the frequency domain, we obtain the following set of equations: 
\begin{widetext}
\begin{align}
    A_1 u_{(4)} + A_2 u_{(4)}'+ A_3 u_{(4)}''+ A_4 u_{(4)}'''&=0\\
    B_1 u_{(2)}+ B_2 u_{(2)}'+ B_3 u_{(2)}''+ B_4 u_{(1)}+ B_5 u_{(1)}'+ B_6 u_{(1)}''+ B_7 u_{(1)}'''+ B_8 u_{(3)}+ B_9 u_{(3)}'+ B_{10} u_{(3)}''&=0\\
    C_1 u_{(2)}+ C_2 u_{(2)}'+ C_3 u_{(2)}''+C_4 u_{(1)}+ C_5 u_{(1)}'+C_6 u_{(3)}+C_7 u_{(3)}'+C_8 u_{(3)}''&=0
\end{align}
These consist of a single, third-order axial equation for the wavefunction $u_{(4)}(r)$, and two equations for the polar wavefunctions $u_{(1)}(r)$, $u_{(2)}(r)$ and $u_{(3)}(r)$. The system can then be closed using the Lorenz condition given by Eq.~\eqref{eq:lorenzcond}. The coefficient of the above equations are listed in the following:
\begin{align}
 A_1=& -4 M r^{5/2}(\lambda +r^2 \mu^2)\omega +2 r^{7/2}\omega(\lambda +r^2 (\mu^2-\omega^2)+2 i \sqrt{2} M^{3/2} r (-2 \lambda +r^2 (6\mu^2-5\omega^2))\nonumber\\
 &+4 i \sqrt{2}M^{5/2}(\lambda -3 r^2 \mu^2)+ i\sqrt{2 M}r^2 (\lambda + 3r^2 (-\mu^2+\omega^2))+2 r^{11/2}  \omega f  \omega_{\rm pl}^{\chi \, 2})\, ,\\
    A_2=& -2 f \sqrt{M} r (-2\sqrt{2} i M^2 +2\sqrt{M} r^{5/2}\omega +\sqrt{2} r^2 i (\lambda +r^2 (\mu^2-\omega^2)+r^2\omega_{\rm pl}^{D \, 2})\nonumber\\
    &-\sqrt{2} i M r (-1+2\lambda+2 r^2 \mu^2+2r^2\omega_{\rm pl}^{\chi \, 2}))\, ,\\
 A_3=&f^2 r^3 (2\sqrt{2} i M^{3/2} +3 i \sqrt{2 M} r-2 r^{5/2}\omega) \, , \\
    A_4= &2\sqrt{2}i f^3 \sqrt{M} r^5 \, ,\\
    B_1=&6\sqrt{2}(M r)^{3/2}\lambda-4\sqrt{2}M^{5/2}\sqrt{r}\lambda+4\sqrt{2}(M r)^{5/2}\mu^2-8 i M^3 r \omega+18 i M^2 r^2\omega- i M r^3 (3+2\lambda+2 r^2 \mu^2)\omega\nonumber\\
    &-2\sqrt{2}M^{3/2}r^{7/2}(\mu^2-4\omega^2)-\sqrt{2 M}r^{5/2}(2\lambda+r^2\omega^2)+ i r^4 \omega (\lambda+r^2(\mu^2-\omega^2))+i f^2 r^6 \omega \omega_{\rm pl}^{\chi \, 2} \, ,\\
 B_2=&-f\sqrt{M}r^{5/2}(2\sqrt{2}M(\lambda+r^2\mu^2)+10 i M^{3/2}\sqrt{r}\omega-i \sqrt{M}r^{3/2}\omega-\sqrt{2}r(\lambda+r^2(\mu^2-2\omega^2))) \, , \\
 B_3=&2 i f^2 M r^5 \omega \, , \\
   B_4= &-16 M^2 r \lambda +4 M^3 (3\lambda-r^2\mu^2)-i\sqrt{2M}r^{7/2}(-1+\lambda+r^2\mu^2)\omega+2\sqrt{2}i M^{3/2}r^{5/2}(\lambda+r^2\mu^2)\omega+r^5\omega^2\nonumber\\
   &+M r^2(5\lambda-2\sqrt{2 M r}i\omega+r^2(\mu^2-2\omega^2))-f^3 r^5\omega_{\rm pl}^{\chi \, 2} \, ,\\
    B_5=&f r^2 (4M^2(\lambda+r^2\mu^2)-2 M r (\lambda+2r^2\mu^2)-\sqrt{2 M}i r^{5/2}\omega-r^4\omega^2+f^2r^4\omega_{\rm pl}^{\chi \, 2}) \, ,\\
    B_6=&f^2\sqrt{M}r^3(2M^{3/2}-\sqrt{M} r+2\sqrt{2}i r^{5/2}\omega) \, , \\
 B_7=&2f^3 M r^5 \, ,\\
    B_8=&-i f\sqrt{M}r^2\omega(2M^{3/2}-3\sqrt{M}r+i\sqrt{2}r^{5/2}\omega) \, , \\
 B_9=&f^2r^{5/2}(-2\sqrt{2}M^{3/2}+\sqrt{2M}r-2iMr^{3/2}\omega-i\omega r^{5/2} \, , \\
 B_{10}=&\sqrt{2M}r^{3/2}(8M^3+6Mr^2-r^3-12M^{3/2}\sqrt{r M}r \, ,\\
    C_1=&f r \lambda (2\sqrt{2}M^{3/2}+2i\omega r^{5/2}+\sqrt{2M}r(-1+2r^2\omega_{\rm pl}^{\chi \, 2})) \, ,\\
 C_2=&r(-4\sqrt{2}\lambda M^{3/2}r+\sqrt{2 M}\lambda r^2+4\sqrt{2}M^{5/2}\lambda+4 i M r^{5/2}\lambda\omega-2 i r^{7/2}\lambda\omega) \, ,\\
 C_3=&-2\sqrt{2M}r^2(\lambda r(-4M+r)+4M^2\lambda) \, ,\\
    C_4=&r^2\lambda(\omega(i\sqrt{2M}(6M-r)+2 r^{5/2}\omega)-2 f r^{5/2}\omega_{\rm pl}^{\chi \, 2}) \, ,\\
 C_5=&-2\sqrt{2M}i f r^4\lambda\omega \, ,\\
    C_6=&r^2(-12\sqrt{2}M^{5/2}\mu^2+2\sqrt{2}M^{3/2} r (6\mu^2-5\omega^2)+3\sqrt{2M}r^2(-\mu^2+\omega^2)+4 i M r^{5/2}\omega(\mu^2+\omega_{\rm pl}^{\chi \, 2})\nonumber\\&-2 i r^{7/2}\omega(\mu^2+\omega_{\rm pl}^{\chi \, 2}-\omega^2))\, , \\
    C_7=&2f\sqrt{M}r(2\sqrt{2}M^2+2i\sqrt{M}r^{5/2}\omega-\sqrt{2}r^4(\mu^2+\omega_{\rm pl}^{\chi \, 2}-\omega^2)+\sqrt{2}M r(-1+2r^2(\mu^2+\omega_{\rm pl}^{\chi \, 2})) \, ,\\
    C_8=&f^2 r^3(2\sqrt{2}M^{3/2}+3\sqrt{2M}r+2 i r^{5/2}\omega) \, ,
\end{align}
where for simplicity we defined $\lambda=l(l+1)$.

\end{widetext}

\end{document}